\documentclass[reprint,amsmath,amssymb,aip,jap]{revtex4-1}
\usepackage{graphicx}
\usepackage[T1]{fontenc}
\usepackage[utf8x]{inputenc}
\usepackage{lmodern} 
\usepackage{multirow}
\usepackage{upgreek}
\usepackage{epstopdf}
\usepackage{booktabs}
\usepackage[separate-uncertainty = true, output-decimal-marker={.}]{siunitx}
\usepackage{tabularx}

\newcolumntype{P}[1]{>{\centering\arraybackslash}p{#1}}

\begin{document}


\title{Experimental prototype of a spin-wave majority gate} 


\author{T. Fischer}
\email[Electronic mail: ]{tfischer@physik.uni-kl.de}
\affiliation{Fachbereich Physik and Landesforschungszentrum OPTIMAS, Technische Universit\"at Kaiserslautern, 67663 Kaiserslautern, Germany}
\affiliation{Graduate School Materials Science in Mainz, 55128 Mainz, Germany}

\author{M. Kewenig}
\affiliation{Fachbereich Physik and Landesforschungszentrum OPTIMAS, Technische Universit\"at Kaiserslautern, 67663 Kaiserslautern, Germany}

\author{D. A. Bozhko}
\affiliation{Fachbereich Physik and Landesforschungszentrum OPTIMAS, Technische Universit\"at Kaiserslautern, 67663 Kaiserslautern, Germany}
\affiliation{Graduate School Materials Science in Mainz, 55128 Mainz, Germany}

\author{A. A. Serga}
\affiliation{Fachbereich Physik and Landesforschungszentrum OPTIMAS, Technische Universit\"at Kaiserslautern, 67663 Kaiserslautern, Germany}

\author{I. I. Syvorotka}
\affiliation{Department of Crystal Physics and Technology, Scientific Research Company Carat, 79031 Lviv,Ukraine}

\author{F. Ciubotaru}
\affiliation{imec, B-3001 Leuven, Belgium}
\affiliation{KU Leuven, Departement Electrotechniek (ESAT), B-3001 Leuven, Belgium}

\author{C. Adelmann}
\affiliation{imec, B-3001 Leuven, Belgium}

\author{B. Hillebrands}
\affiliation{Fachbereich Physik and Landesforschungszentrum OPTIMAS, Technische Universit\"at Kaiserslautern, 67663 Kaiserslautern, Germany}

\author{A. V. Chumak}
\affiliation{Fachbereich Physik and Landesforschungszentrum OPTIMAS, Technische Universit\"at Kaiserslautern, 67663 Kaiserslautern, Germany}


\date{\today}

\begin{abstract}
Featuring low heat dissipation, devices based on spin-wave logic gates promise to comply with increasing future requirements in information processing. In this work, we present the experimental realization of a majority gate based on the interference of spin waves in an Yttrium-Iron-Garnet-based waveguiding structure. This logic device features a three-input combiner with the logic information encoded in the phase of the spin waves. We show that the phase of the output signal represents the majority of the phase of the input signals. A switching time of about $\SI{10}{ns}$ in the prototype device provides evidence for the ability of sub-nanosecond data processing in future down-scaled devices.
\end{abstract}


\maketitle 


The scaling of conventional CMOS-based nanoelectronics is expected to become increasingly intrinsically limited in the next decade. Therefore, novel beyond-CMOS devices are being actively developed as a complement to expand functionally in future nanoelectronic technology nodes \cite{Nikonov.2013}. In particular, the field of magnonics \cite{Hertel.2004,Khitun.2005,Kruglyak.2006,Schneider.2008,Chumak.2014,Klingler.2014} (see also reviews \citenum{Serga.2010,Khitun.2010,Kruglyak.2010,Lenk.2011,Chumak.2015}) which utilizes the fundamental excitations of a magnetic system - spin waves \cite{Bloch.1930} and their quanta - magnons \cite{Holstein.1940} as data carriers, provides promising approaches to overcome crucial limitations of CMOS since they may provide ultralow power operation as well as nonvolatility \cite{Khitun.2010,Dutta.2015b,Chumak.2015}. Magnonic devices are especially amenable to building majority gates \cite{Khitun.2011,Klingler.2014,Klingler.2015,Radu.2015,Ganzhorn.2016} with excellent scaling potential leading to an improved circuit efficiency. Hence, majority gates can be considered to be key devices in a novel approach to circuit design with strongly improved area and power scaling behavior \cite{Amaru.2014}.

Spin waves cover characteristic frequencies in the GHz regime and their wavelength can easily be reduced down to the nanometer range \cite{Sandweg.2011,Chumak.2015}. Furthermore, their dispersion relation is highly versatile depending on material parameters as well as magnetization and field configuration \cite{Serga.2010} making them usable in a wide range of devices \cite{Hertel.2004,Khitun.2005,Kruglyak.2006,Schneider.2008,Chumak.2010,Kruglyak.2010,Inoue.2011,Chumak.2014,Klingler.2014,Schultheiss.2014,Nikitin.2015}. In this context, majority gates are of special interest since a simple spin-wave combiner substitutes several tens of transistors, and three majority gates suffice for creating a full-adder \cite{Martinez.2010}. Multi-frequency operation allows for parallel data processing \cite{Khitun.2012}.

In this work, we present the experimental realization and investigation of a prototype of a spin-wave majority gate, whose functionality and performance on the microscopic scale have been investigated in numerical simulations \cite{Klingler.2014,Klingler.2015}. 
The investigated majority gate has three inputs and one output \cite{Khitun.2010}. The output state represents the majority of the input states, which is illustrated in Table \ref{truthtable}. In the numerical investigations as well as in our experiments, the information is encoded in the phase of the spin waves. Here, a logic `0' corresponds to a certain phase $\phi^{(0)}$, whereas a logic `1' is represented by a phase of $\phi^{(1)}=\phi^{(0)}+\pi$. Without loss of generality, the respective phases can be redefined to be $\phi^{(0)}=0$ for a logic `0' and $\phi^{(1)}=\pi$ for a logic `1'.
\begin{table}%

\begin{tabular}{P{0.7cm}P{0.7cm}P{0.7cm}cc}
\hline
\multicolumn{3}{c}{Input phases}&\multirow{2}{*}{\parbox{2.2cm}{Logic input \\ state}}&\multirow{2}{*}{\parbox{2.2cm}{Logic output \\ state}}\\
i1&i2&i3&&\\
\hline
0&0&0&000&0\\
0&0&$\pi$&001&0\\
0&$\pi$&0&010&0\\
$\pi$&0&0&100&0\\
$\pi$&0&$\pi$&101&1\\
$\pi$&$\pi$&0&110&1\\
0&$\pi$&$\pi$&011&1\\
$\pi$&$\pi$&$\pi$&111&1\\
\hline
\end{tabular}
\caption{Truth table of the majority operation. In the case of a spin-wave majority gate, the logic state is encoded in the phase of the waves.}
\label{truthtable}
\renewcommand*{\arraystretch}{1}
\end{table}
\begin{figure*}[bhpt]
\includegraphics[width=1\textwidth]{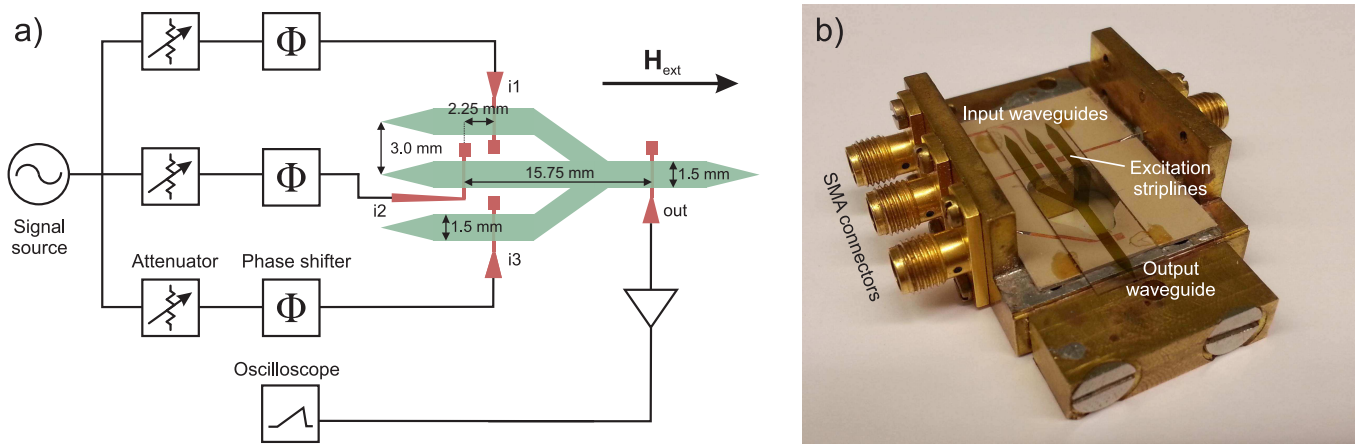}%
\caption{\label{mwsetup} Sketch of the microwave setup used in the experiments (a) and photograph of the device under test (b). The input signals diverge from a single microwave source and are conditioned in each channel with respect to amplitude and phase. The signals are then sent to the shortcut copper striplines (red) evoking an Oersted field which excites spin precession. Spin waves propagate through the input waveguides (width $w_g=\SI{1.5}{mm}$) and the spin-wave combiner towards the output waveguide (green). The output signal is detected employing a fast oscilloscope.}%
\end{figure*}

In each of the three input spin-wave waveguides spin waves are excited inductively by RF currents sent through copper striplines. These striplines have been structured by means of wet-chemical etching from an RF laminate (\textit{Rogers RO4000\textregistered}) and feature a width of $w_a=\SI{75}{\um}$. All spin-wave waveguides have a width of $w_g=\SI{1.5}{mm}$. After propagating through the input waveguides, spin waves enter the spin-wave combiner in which the waveguides merge into a single waveguide, subsequently leading to a superposition of the excited waves. The propagation of waves through the skews into the combiner and the output waveguide is expected to feature comparably large losses. However, it has been shown that propagation of spin waves through bended structures is possible \cite{Clausen.2011}.
\begin{figure}[]
\includegraphics[width=0.48\textwidth]{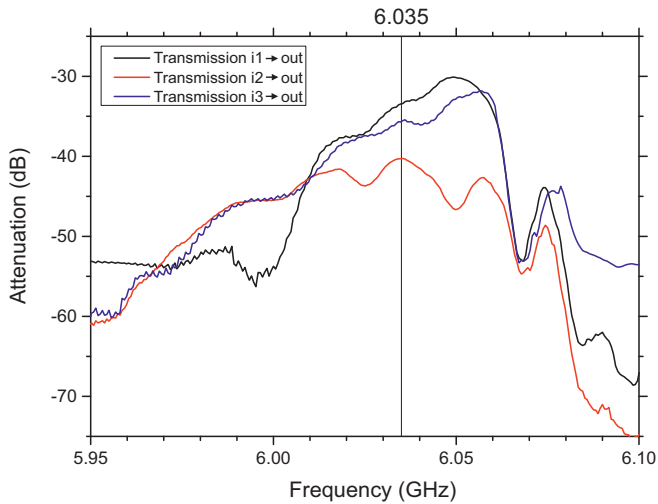}%
\caption{\label{transmission} Transmission spectra of the respective input channels towards the output channel. The characteristic backward volume dispersion behavior is clearly visible. Differences in the attenuation are the result of various effects such as different excitation efficiency, electromagnetic coupling, and different properties of the microwave components.}%
\end{figure}
The spin-waves in the output waveguide again generate an electrical signal in a copper stripline which can be detected electrically by using a rectifying diode and an oscilloscope, or an oscilloscope with a sufficient sampling rate to directly map the spin-wave signal, respectively. The actual waveguiding structure is depicted in green in Fig. \ref{mwsetup}a together with the microwave setup employed for signal conditioning and the excitation of spin waves. The RF current for the excitation of spin waves which is provided by a single microwave generator (\textit{Agilent E8257D}) is split into three input signals. The input channels each feature an adjustable phase shifter in order to encode the information into the phase of the signals exciting the spin waves. Attenuators are used to level out the different transmission characteristics of the equipment and the different spin-wave excitation efficiencies of the respective input striplines. The oscilloscope allows for the measurement of the resulting spin-wave signal in the output channel.
A photograph of the device investigated in this work together with the underlying excitation structure and microwave connectors is shown in Fig. \ref{mwsetup}b. 

The magnetic structure has been fabricated by means of photolithography and wet-chemical etching from a YIG (Yttrium Iron Garnet, $\text{Y}_{3}\text{Fe}_5\text{O}_{12}$) film with a thickness of $\SI{5.4}{\um}$ grown on a Gallium Gadolinium Garnet substrate by means of liquid phase epitaxy \cite{Syvorotka.2010,Syvorotka.2015}. Before undergoing the patterning process, the linewidth of the ferromagnetic resonance of the film has been determined to $\mu_0 \Delta H_0=\SI{0.062}{mT}$. Since this configuration allows for mode selection in the output waveguide in the case of microstructures \cite{Klingler.2014}, the device is operated in the regime of backward volume spin waves \cite{Serga.2010} with the external field applied parallel to the waveguides. However, we have also tested the device in the configuration of magnetostatic surface spin waves with the field applied perpendicularly to the waveguides. Although the transmission spectra differ, the operational characteristics appear to be qualitatively the same.
In order to determine the parameters for an efficient operation of the device, spin-wave transmission through each of the channels towards the output is measured employing a Vector Network Analyzer (\textit{Anritsu MS46322A}). 
In order to avoid nonlinear effects, the input power is kept at a level of \SI{-3}{dBm}. The corresponding transmission characteristics for each input to the output are presented in Fig. \ref{transmission}. For all three spectra, the backward volume transmission characteristic is visible with the FMR frequency lying at about $f_{\text{FMR}}\approx\SI{6.06}{GHz}$. The transmission highly depends on various factors, such as the excitation efficiency at each input stripline, reflections in the bends in the combiner section of the gate, and the electromagnetic coupling between the respective input stripline and the output stripline, resulting in variations of the transmission efficiency for the different channels. Mainly, this is attributed to the anisotropic character of spin waves in in-plane magnetized films: The transmission through the combiner requires a modification of the wave vector in the skews. 
For an externally applied field of $\mu_0 H_\text{ext}=\SI{142.9}{mT}$ we select an operational spin-wave carrier frequency of $f_\text{c}=\SI{6.035}{GHz}$ which is used in the subsequent experiments. This is justified in order to minimize distortions of the output signal. Especially in the case of pulsed excitation, a constant transmission over the entire frequency spectrum of the pulse is desirable. Nevertheless, our studies have shown that the choice of the working point is not critical and the majority gate works well in a wide range of frequencies.
For a reliable operation of the device, equality of the signal amplitudes in each channel has to be ensured. Therefore, the attenuators in each channel are adjusted accordingly. As mentioned above, the information is encoded in the phase of the spin waves. Hence, the desired phase shifts of $0$ and $\pi$ for each signal have to be defined in consistency with the phase shifts of the remaining channels. This requirement is satisfied by the following procedure: A signal whose phase is defined as $0$ is directly provided to the input i2 serving as the reference channel. Subsequently, a signal with calibrated amplitude is applied to one of the remaining channels. By adjusting the phase shift and, thus, maximizing and minimizing the output signal, respectively, phase shifts of $0$ and $\pi$ can be assigned to the signal applied to this channel.
\begin{figure}[]
\includegraphics[width=0.48\textwidth]{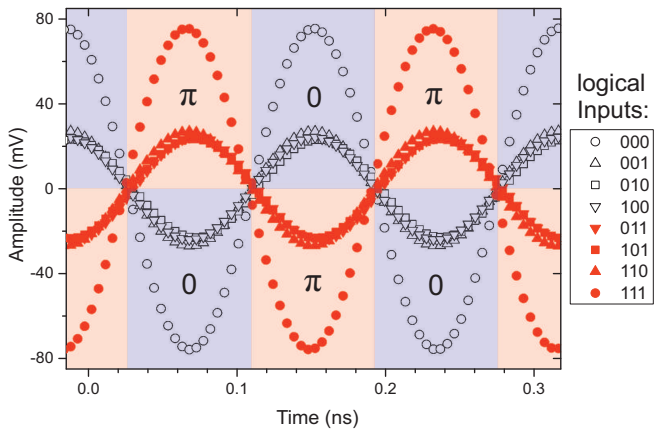}%
\caption{\label{fastoscilloscope} Output signal observed with an oscilloscope with high sampling rate allowing for the direct mapping of the spin-wave amplitude. The corresponding experimental setup is depicted in Fig. \ref{mwsetup}a. The dependence of the output phase on the majority of the input phases is clearly visible.}%
\end{figure}
\begin{figure*}[ht]
\includegraphics[width=1\textwidth]{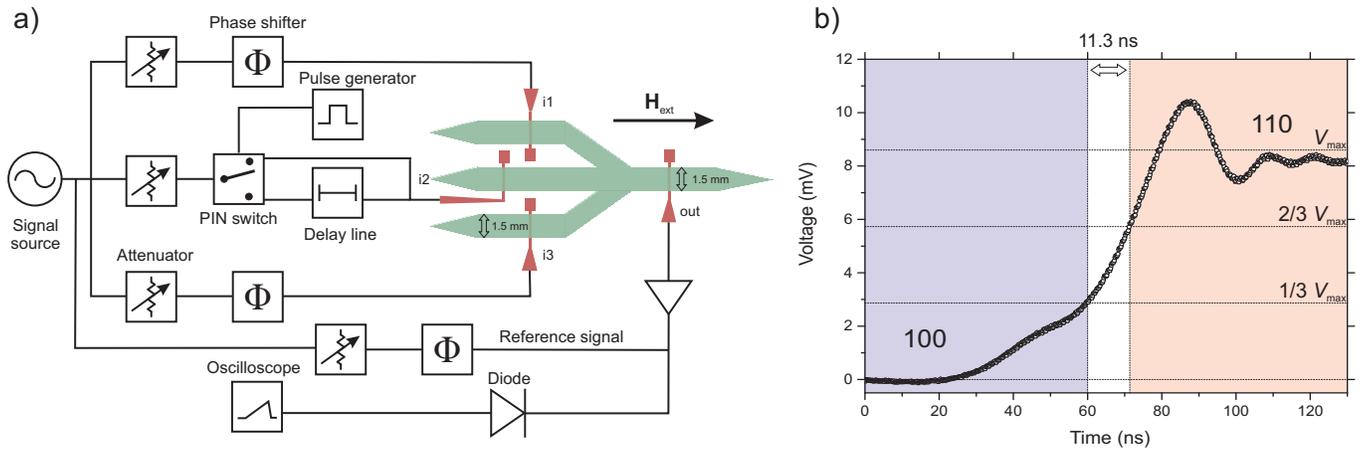}%
\caption{\label{rise} (a) Experimental setup used for the determination of the switching speed of the device. A fast microwave switch and a delay line enable a periodic modulation of the phase if the input signal at i2 hence resulting in a periodic toggling between the logic input states `100' and `110'. (b) Switching event of the output signal recorded with a rectifying diode and an oscilloscope. The signal exhibits a rise time of $t_{\text{rise}}=\SI{11.3}{\ns}$.}%
\end{figure*}

In the following, we will present and discuss the results obtained from our measurements on the logic operation of the device. First, for the detection and mapping of the output signal, a fast oscilloscope with a sampling rate of $\SI{40}{GSa/s}$ (\textit{Agilent infiniium DSO80804B}) is used. This allows for a direct detection of the spin-wave signal in the output waveguide which is picked up by the output stripline. 
For this, all possible input states have been applied to the device and the respective output signal for each of these states has been recorded. The results of these measurements are presented in Fig. \ref{fastoscilloscope}. 
The quantity most relevant to the output signal is its phase, since this parameter serves as the carrier of information. As can be seen, for instance, from the output signals for the states `100' (red downward triangles) and `011' (black squares), respectively, Fig. \ref{fastoscilloscope} reveals that the phase of the output signal shifts by $\pi$ if all input signals are shifted by $\pi$. Defining the corresponding phases of the output signal as logic `0' and `1', respectively, a reliable operation of the device is possible with the phase of the output signal representing the majority of the phase of the input signals. 
The data also show that output amplitude depends on the input state, as expected for interference based wave computing. For instance, in case of the input state ‘111’, the output amplitude is $U_\text{a}\approx\SI{75}{mV}$ whereas for the input state ‘110’ it amounts to $U_\text{a}\approx\SI{25}{mV}$. As a consequence, spin-wave based majority gates cannot be cascaded directly and more complex (clocked) interconnecting schemes are required with the added benefit of nonvolatility \cite{Dutta.2015b}. Alternative approaches might be provided by the concept of parametric amplification \cite{Bracher.2016} or nonlinear magnon phenomena \cite{Chumak.2014}.

In addition to the direct mapping of the spin-wave signal in the output channel, we are interested in the switching speed of the device. The clock rate of potential computing devices based on the majority gate is limited by the time required for the gate to switch its logic output state. In order to investigate this experimentally, we changed the input signal instantly between the logic states `100' and `110'. The corresponding microwave setup is shown in Fig. \ref{rise}a. The input state is switched by inserting a fast microwave switch into the input channel i2. By applying pulses to the switch (rise time $t_\text{rise} < \SI{2}{\ns}$), the corresponding input signal either is directly applied to the excitation stripline or first passes a delay line resulting in a phase shift of $\pi$. Since the output signal only changes its phase and not its amplitude, we combine it with a reference signal with phase $\phi_\text{ref}=\pi$. The envelope of the interference signal is measured employing a diode which allows for analyzing the resulting intensity of the combined signals.
The results of this measurement are shown in Fig. \ref{rise}b. We define logic voltage levels assuming threshold voltage values of $1/3\times V_{\text{max}}$ and $2/3\times V_{\text{max}}$ for logic `0' and logic `1', respectively.
Taking into account this definition, the rise time of the output signal switching from logic `0' to logic `1' amounts to $t_\text{rise}=\SI{11.3}{ns}$ resulting in a clock frequency of a potential device of $t_\text{rise}^{-1}=f_\text{clock}=\SI{88.5}{MHz}$. Since the clock frequency is related to the propagation length of the waves through gate structure, we expect the rise time to show good scaling behavior, and switching times below $\SI{1}{ns}$ should be feasible by miniaturizing the device.

It can be summarized that in this work we succeeded in the experimental realization of a majority gate based on the interference of spin waves. Here, the phase of the output signal is defined by the majority of the phases of the input signals. Recent progress in the miniaturization of YIG structures \cite{Hahn.2014,Onbasli.2014,Dubs.2016} as well as techniques for a phase control by electric fields as described in the work by Ustinov \textit{et al.} \cite{Ustinov.2014} or by spin-polarized currents \cite{Chen.2015c} provide promising drafts for the required advances towards applications. Due to their different properties, alternative materials such as Heusler compounds \cite{Kubota.2009,Trudel.2010} might constitute an interesting option for the realization of a spin-wave majority gate on the micro- and nanoscale.


%
%

%

\begin{acknowledgments}
T.F. and D.A.B. have been supported by a fellowship through the Excellence Initiative (DFG/GSC 266). A.V.C. gratefully acknowledges financial support by the European Union within the ERC Starting Grant 678309 MagnonCircuits.
This work was also supported by the European Union within the EU-FET grant InSpin 612759.
\end{acknowledgments}

\bibliography{Bibliography}


\end{document}